\documentclass[twocolumn,showpacs,preprintnumbers,amsmath,amssymb]{revtex4}

\usepackage{bm}

\newcommand{\bv}{\bar{\varphi}}

\begin{document}

\title{Cosmological constant and quantum gravitational corrections to the running fine structure constant}

\author{David J. Toms}
\homepage{http://www.staff.ncl.ac.uk/d.j.toms}
\email{d.j.toms@newcastle.ac.uk}
\affiliation{%
School of Mathematics and Statistics,
Newcastle University,
Newcastle upon Tyne, U.K. NE1 7RU}

\date{\today}

\begin{abstract}
The quantum gravitational contribution to the renormalization group behavior of the electric charge in Einstein-Maxwell theory with a cosmological constant is considered. Quantum gravity is shown to lead to a contribution to the running charge not present when the cosmological constant vanishes. This re-opens the possibility, suggested by Robinson and Wilczek, of altering the scaling behaviour of gauge theories at high energies although our result differs. We show the possibility of an ultraviolet fixed point that is linked directly to the cosmological constant.
\end{abstract}


\maketitle

The behaviour of the coupling constants in quantum field theory at different energy, or length, scales is governed by the Callan-Symanzik~\cite{Callan,Symanzik}, or renormalization group equations. For a theory with a single coupling constant $g$, if we scale the energy $E$ by $E\rightarrow e^{t}E$ the running coupling constant $g(t)$ usually obeys the equation
\begin{equation}\label{CS}
\frac{dg(t)}{dt}=\beta(g(t))
\end{equation}
for some function $\beta(g)$, called the renormalization group $\beta$-function. This equation governs the ultraviolet (infrared) behaviour of the theory obtained from $t\rightarrow\infty\ (t\rightarrow-\infty)$. Zeroes of the $\beta$-function are called fixed points. In particular, if $g(t)\rightarrow0$ as $t\rightarrow\infty$ the theory is said to be asymptotically free, meaning that the coupling constant becomes weaker at high energy (short distance). For a review of renormalization group terminology, see \cite{Gross75} or \cite{Weinberg} for example.

One of the great calculations of modern quantum field theory demonstrates that Yang-Mills gauge theories can exhibit asymptotic freedom \cite{GrossandWilczek,Politzer}. Such theories lie at the heart of modern particle physics as encoded in the standard model. However, if gravity is considered classic papers of `t~Hooft and Veltman \cite{tHooftVeltman} and Deser, van Nieuwenhuizen and Tsao \cite{DeservanN1,DeservanN2,DeservanN3,DeserTsaovanN} demonstrate that the coupling of matter fields to Einstein's theory of gravity renders the full theory non-renormalizable. This suggests that Einstein's theory of gravity may not be fundamental in the same sense as Yang-Mills theory. However, it still makes sense to consider the perturbative quantization of Einstein's theory within the effective field theory framework as emphasized in the important papers of Donoghue \cite{Donoghue}.

With this effective field theory viewpoint in mind, Robinson and Wilczek~\cite{RobWilczek} claimed that quantum gravitational corrections to the renormalization group functions tended to make all theories asymptotically free at high enough energies. This holds even for theories like QED that do not exhibit asymptotic freedom in the absence of gravity. If true, there are possible phenomenological consequences \cite{Gogoladze}. In addition, the results of \cite{RobWilczek} have been used to justify~\cite{Huang} the weak gravity effect~\cite{ArkaniHamed}. With these motivations it is important to examine the conclusions of Robinson and Wilczek in other independent calculations.

Doubt on the validity of the results of Robinson and Wilczek was cast by Pietrykowski~\cite{Piet} who showed for Einstein-Maxwell theory that their result was gauge dependent, and that a different choice of gauge leads to a null result. This was demonstrated in a gauge condition independent way in Ref.~\cite{TomsPRD}. The calculation of Robinson and Wilczek involves quadratically divergent integrals that get regularized to zero if dimensional regularization is used. Standard renormalization group arguments of `t~Hooft~\cite{tHooftRG} show that only logarithmic divergences contribute to renormalization group functions affecting running coupling constants. A very important contribution~\cite{Ebertetal} (see also \cite{RodigastPhD}) to the debate over the quantum gravity corrections to the running coupling constants demonstrates conclusively that all of the quadratic divergences that are present in the theory cancel in a cut-off regularization scheme, and that the logarithmic divergences that are present agree with those found using dimensional regularization; they too are absent, and there is no purely quantum gravity contribution to the running gauge coupling constants. (An earlier string calculation~\cite{Kiritsis} showed that there was no gravitational correction to the gauge coupling constants.)

The calculations of \cite{RobWilczek,Piet,TomsPRD,Ebertetal,RodigastPhD} are all for Einstein gravity without a cosmological constant. What we will show in this paper is that the quantization of Einstein-Maxwell theory with a cosmological constant present does lead to a running of the electric charge, and by extrapolation, to a running of the Yang-Mills gauge coupling constant. However the scaling behaviour of the theory is different from that predicted originally in \cite{RobWilczek}, and this modified behaviour could lead to a new fixed point that is not found in the absence of gravity. The quantization of Einstein-Maxwell theory with and without a cosmological constant was first given in \cite{DeservanN2}; however it is worth a reexamination for reasons we now discuss.

Our method of calculation will follow our earlier paper~\cite{TomsPRD} and use the gauge-invariant background-field method and dimensional regularization. The calculation is of necessity off-shell since the classical equations of motion are not satisfied with the background spacetime and electromagnetic field that are needed. This is true for two reasons. In the first place we are expanding about flat Minkowski spacetime that is not a solution to the Einstein equations with a cosmological constant included. The second reason is that even if the cosmological constant vanishes we must keep the electromagnetic field non-zero if we are to calculate a possible contribution to the $\beta$-function for the running charge, and this does not solve the Einstein field equations either. Unless special care is taken the results of the calculation can be gauge dependent as well as dependent on the gauge conditions that must be imposed. One way to ensure that this does not happen is to use the background-field method of Vilkovisky~\cite{Vilkovisky} and DeWitt~\cite{DeWitt}.

To one-loop order the effective action is \cite{HKLT}
\begin{eqnarray}
\Gamma\lbrack\bv\rbrack&=&S\lbrack\bv\rbrack - {\rm ln\,det}\;Q_{\alpha\beta}\lbrack\bv\rbrack\nonumber\\
&&\hspace{-24pt}+\frac{1}{2}\lim_{\xi\rightarrow0} {\rm ln\,det}\left\lbrace \nabla^i\nabla_j S\lbrack\bv\rbrack +\frac{1}{2\xi}K^{i}_{\alpha}\lbrack\bv\rbrack K^{\alpha}_{j}\lbrack\bv\rbrack\right\rbrace\;, \label{eq4}
\end{eqnarray}
where
\begin{eqnarray}
\nabla_i\nabla_j S\lbrack\bv\rbrack&=&S_{,ij}\lbrack\bv\rbrack-\Gamma^{k}_{ij}\lbrack\bv\rbrack S_{,k}\lbrack\bv\rbrack\;,\label{eq5}\\
Q_{\alpha\beta}\lbrack\bv\rbrack&=&\frac{\delta\chi_\alpha}{\delta\epsilon^\beta}\;.\label{eq6}
\end{eqnarray}
We use DeWitt's condensed notation~\cite{DeWittdynamical} here with $\varphi^i$ the complete set of fields. Here $K^{i}_{\alpha}$ are the generators of gauge transformations defined by $\delta\varphi^i=K^{i}_{\alpha}\delta\epsilon^\alpha$, with $\delta\epsilon^\alpha$ the infinitesimal parameters of the gauge transformation. $\chi_\alpha$ denotes the gauge condition and $\Gamma^{k}_{ij}$ is the connection on the space of fields. Provided that we adopt the Landau-DeWitt gauge condition
\begin{equation}\label{eq3}
\chi_\alpha=K_{\alpha i}\lbrack\bv\rbrack(\varphi^i-\bv^i)=0\;,
\end{equation}
$\Gamma^{k}_{ij}$ may be taken to be the Christoffel connection associated with the field space metric. It is worth emphasizing that the background field $\bv^i$ is completely arbitrary here, and that it is not necessary to expand about a classical solution to the equations of motion.  The resulting effective action is completely independent of the choice of gauge condition; the Landau-DeWitt choice (\ref{eq3}) is for ease of computation only. The contribution from $Q_{\alpha\beta}$ is the usual Faddeev-Popov ghost term and, as it is unchanged by the presence of a cosmological constant, may be shown~\cite{TomsPRD} to make no contribution to the part of the effective action responsible for charge renormalization.

The classical action functional for Einstein-Maxwell theory is
\begin{equation}\label{act1}
S=S_G+S_M\;,
\end{equation}
where
\begin{equation}\label{act2}
S_G=-\frac{2}{\kappa^2}\int d^nx|g(x)|^{1/2}(R-2\Lambda)\;,
\end{equation}
is the gravitational Einstein-Hilbert action with $\kappa^2=32\pi G$ and $\Lambda$ the cosmological constant with
\begin{equation}\label{act3}
S_M=\frac{1}{4}\int d^nx|g(x)|^{1/2}F_{\mu\nu}F^{\mu\nu}
\end{equation}
the Maxwell action for the electromagnetic field. We choose the curvature conventions of \cite{MTW} but with a Riemannian metric. We choose $\varphi^i=(g_{\mu\nu}(x),A_\mu(x))$, but this is totally arbitrary; any other choice will lead to the same result since the formalism is independent of the choice of field parameterization.

Because our concern here is only the possibility of charge renormalization as a consequence of quantum gravity corrections, we simplify by choosing the background spacetime to be flat, $g_{\mu\nu}=\delta_{\mu\nu}$, and the background gauge field $\bar{A}_\mu$ to correspond to a constant electromagnetic field $\bar{F}_{\mu\nu}$. We can concentrate on pole terms in the one-loop effective action that involve only $\bar{F}_{\mu\nu}\bar{F}^{\mu\nu}$ since standard background-field methods link the field renormalization factor to that of the charge~\cite{Abbott}.

We can write~\cite{TomsPRD}
\begin{eqnarray}
-\frac{1}{2} {\rm ln\,det}\left\lbrace \nabla^i\nabla_j S\lbrack\bv\rbrack +\frac{1}{2\xi}K^{i}_{\alpha}\lbrack\bv\rbrack K^{\alpha}_{j}\lbrack\bv\rbrack\right\rbrace&&\nonumber\\ &&\hspace{-108pt}=\ln\int\lbrack d\eta\rbrack\,e^{-S_q}\;,\label{eq20}
\end{eqnarray}
where
\begin{equation}\label{eq21}
S_q=\frac{1}{2}\eta^i\eta^j\left(S_{,ij}-\Gamma^{k}_{ij}S_{,k} +\frac{1}{2\xi}K_{\alpha\,i}K^{\alpha}_{j}\right)\;.
\end{equation}
Our interest is in evaluating terms in the effective action that are of quadratic order in the background electromagnetic field. We separate $S_q$ in (\ref{eq21}) into three terms,
\begin{equation}\label{eq22a}
S_q=S_0+S_1+S_2\;,
\end{equation}
where the subscript $0,1,2$ counts the power of the background field $\bar{A}_\mu$ that occurs. After some calculation it is found that
\begin{eqnarray}
S_0&=&\int d^nx\Big\lbrace-\frac{1}{2}h^{\mu\nu}\Box h_{\mu\nu}+\frac{1}{4}h\Box h\nonumber\\ &&+\left(\frac{1}{\kappa^2\xi}-1\right)\left(\partial^\mu h_{\mu\nu}-\frac{1}{2}h_{,\nu}\right)^2\nonumber\\
&&-\frac{n\Lambda}{2(n-2)}(h^{\mu\nu}h_{\mu\nu}-\frac{1}{2}h^2)-\frac{1}{2}\Lambda a^\mu a_\mu\nonumber\\
&&+\frac{1}{2}a_\mu(-\delta^{\mu\nu}\Box+\partial^\mu\partial^\nu)a_\nu +\frac{1}{4\zeta}(\partial^\mu a_\mu)^2\Big\rbrace,\label{S0}\\
S_1&=&\frac{\kappa}{2}\int d^nx\Big(h\bar{F}^{\mu\nu}-2h^{\mu}{}_{\lambda}\bar{F}^{\lambda\nu}+2h^{\nu}{}_{\lambda}\bar{F}^{\lambda\mu} \Big)a_{\nu,\mu}\nonumber\\
&&+\frac{1}{\kappa\xi}\int d^nx\Big(\partial^\mu h_{\mu\nu}-\frac{1}{2}h_{,\nu}\Big)\bar{F}^{\lambda\nu}a_\lambda,\label{S1}\\
S_2&=&{\kappa^2}\int d^nx\Big\lbrace \frac{1}{4}\bar{F}_{\mu\nu}\bar{F}^{\mu}{}_{\sigma}\left( h^{\nu\lambda}h_{\lambda}^{\sigma}-\frac{1}{2}hh^{\nu\sigma}\right)\nonumber\\
&& -\frac{(4-n)}{32(2-n)}\bar{F}^2\left( h^{\nu\lambda}h_{\lambda}^{\sigma}-\frac{1}{2}hh^{\nu\sigma}\right)\nonumber\\
&&-\frac{1}{32}a_{\mu}a_{\nu}(\delta^{\mu\nu}\bar{F}^2- 4\bar{F}^{\mu\lambda}\bar{F}^{\nu}{}_{\lambda})\nonumber\\
&&+\frac{1}{4}\bar{F}_{\mu\nu}\bar{F}_{\lambda\sigma}h^{\mu\lambda}h^{\nu\sigma}+ \frac{1}{4\xi\kappa^2}\bar{F}_{\mu\lambda}\bar{F}_{\nu}{}^{\lambda}a^\mu a^\nu\Big\rbrace\;.\label{S2}
\end{eqnarray}
We distinguish the parameters used to implement the Landau-DeWitt gauge conditions for gravity and electromagnetism by $\xi$ and $\zeta$ respectively. We abbreviate $\bar{F}_{\mu\nu}\bar{F}^{\mu\nu}=\bar{F}^2$ and use $h=h^{\mu}{}_{\mu}$. $h_{\mu\nu}$ and $a_\mu$ are the fields integrated over in the functional integral, represented by $\eta^i$ in condensed notation in (\ref{eq20},\ref{eq21}). The spacetime dimension $n$ is kept general at this stage. All Greek indices in (\ref{S0}--\ref{S2}) are normal spacetime labels.

From the $S_0$ term, that determines the propagators, it can be seen that the cosmological constant acts like a mass term for the photon; the origin of this is the field space connection that is absent in the standard background-field approach. Part of the $\Lambda h^2$ term has a similar origin and interpretation. The photon and graviton propagators that follow from $S_0$ in (\ref{S0}) are
\begin{equation}
\left\lbrace\begin{array}{c}
G_{\mu\nu}(x,x')\\
G_{\rho\sigma\lambda\tau}(x,x')\\ \end{array}\right\rbrace =\int\frac{d^np}{(2\pi)^n}\,e^{ip\cdot(x-x')} \left\lbrace\begin{array}{c}G_{\mu\nu}(p)\\
 G_{\rho\sigma\lambda\tau}(p)\\
 \end{array}\right\rbrace\;,\label{photon1}
\end{equation}
where
\begin{equation}\label{photon2}
G_{\mu\nu}(p)=\frac{\delta_{\mu\nu}}{(p^2-\Lambda)}+\frac{(2\zeta-1)\,p_\mu p_\nu}{(p^2-\Lambda) (p^2-2\zeta\Lambda)}\;,
\end{equation}
and
\begin{eqnarray}\label{propgrav}
G_{\rho\sigma\lambda\tau}(p) &=&\frac{\delta_{\rho\lambda}\delta_{\sigma\tau}+\delta_{\rho\tau}\delta_{\sigma\lambda}-\frac{2}{n-2}\delta_{\rho\sigma}\delta_{\lambda\tau}}{2\left(p^2+\frac{n\Lambda}{2-n}\right)}\\
&&\hspace{-48pt}+\frac{1}{2}(\kappa^2\xi-1)\frac{\delta_{\rho\lambda}p_\sigma p_\tau+\delta_{\rho\tau}p_\sigma p_\lambda+\delta_{\sigma\lambda}p_\rho p_\tau+\delta_{\sigma\tau}p_\rho p_\lambda}{\left(p^2+\frac{n\Lambda}{2-n}\right)\left(p^2+\frac{n\kappa^2\xi\Lambda}{2-n}\right)}\nonumber
\end{eqnarray}

To obtain that part of the effective action quadratic in $\bar{A}_\mu$ resulting from the gauge fields $h_{\mu\nu}$ and $a_\mu$ we expand the functional integral representation (\ref{eq20}) in powers of $\bar{A}_\mu$ up to second order. (The first order term vanishes.) We have
\begin{equation}\label{gamma2}
\Gamma_2=\langle S_2\rangle-\frac{1}{2}\langle S_1^2\rangle
\end{equation}
where $\langle\cdots\rangle$ means to evaluate the enclosed expression using the Feynman rules and propagators given in (\ref{photon1}--\ref{propgrav}) keeping only one-particle irreducible graphs. We use
\begin{eqnarray}
\langle h_{\mu\nu}(x)h_{\rho\sigma}(x')\rangle&=&G_{\mu\nu\rho\sigma}(x,x')\;,\label{gravprop2}\\
\langle a_{\mu}(x)a_{\nu}(x')\rangle&=&G_{\mu\nu}(x,x')\;.\label{photprop2}
\end{eqnarray}
Standard results of dimensional regularization lead to (letting $n\rightarrow4$ everywhere except in the pole term)
\begin{equation}\label{S2DR}
\langle S_2\rangle\simeq-\frac{3}{4}\left(\kappa^2+\frac{1}{2\xi}\right)\frac{\Lambda}{16\pi^2(n-4)}\int d^4x\bar{F}^2\;.
\end{equation}
(The $\simeq$ in (\ref{S2DR}--\ref{gammadiv}) denotes that this is not the complete expression, merely the part that can contribute to charge renormalization.) We can safely take the limits $\xi\rightarrow0$ and $\zeta\rightarrow0$ in all but the divergent $1/\xi$ term. This term must cancel with a similar term coming from $\langle S_1^2\rangle$ since the formalism guarantees the existence of the $\xi\rightarrow0$ limit.

Unlike the situation for $\langle S_2\rangle$, the evaluation of $\langle S_1^2\rangle$ does not simply involve the coincidence limit of propagators. Because $S_1$ in (\ref{S1}) involves an explicit factor of $1/\xi$, $\langle S_1^2\rangle$ will involve factors of $1/\xi,1/\xi^2$ as well as terms that remain finite as $\xi\rightarrow0$. It can be shown that all of the $1/\xi^2$ dependence cancels and we are left with (again taking $n\rightarrow4$ everywhere except in the pole term)
\begin{equation}\label{S1eval}
\langle S_1^2\rangle\simeq-\left(\frac{9\kappa^2}{4}+\frac{3}{4\xi}\right) \frac{\Lambda}{16\pi^2(n-4)}\int d^4x\bar{F}^2.
\end{equation}
By combining (\ref{S2DR}) with (\ref{S1eval}) to evaluate (\ref{gamma2}), it can be observed that all of the $1/\xi$ terms that diverge as $\xi\rightarrow0$ cancel to leave us with
\begin{equation}\label{gammadiv}
\Gamma_2\simeq\frac{3\kappa^2\Lambda}{128\pi^2(n-4)}\int d^4x\bar{F}^2\;.
\end{equation}
This is our main result showing that there is a divergence present in the effective action that must be dealt with by a field renormalization. The field renormalization is directly linked to the charge renormalization \cite{Abbott}.

A standard renormalization group analysis \cite{tHooftRG} leads to the equation governing the running electric charge $e(t)$:
\begin{equation}\label{fine1}
\frac{de(t)}{dt}=-\frac{3}{64\pi^2}\kappa^2\Lambda e\;.
\end{equation}
Here $t$ is the parameter that scales the energy $E\rightarrow e^{t}E$ used in (\ref{CS}). It follows that the behaviour of the running electric charge is governed mainly by the sign of the cosmological constant. If $\Lambda>0$ then the quantum gravity correction will lead to asymptotic freedom, whereas a negative cosmological constant will lead to infrared freedom (as $t\rightarrow-\infty$). This conclusion is based on just the quantum gravity contribution, and ignores the contributions from other matter fields. Note that $\kappa$ and $\Lambda$ will satisfy renormalization group equations of their own that must be taken into account when solving (\ref{fine1}).

We can study other possibilities that could arise in more realistic, or more general, theories. Suppose that the renormalization group equation looks like
\begin{equation}\label{fine2}
\frac{dg(t)}{dt}=a\kappa^2\Lambda g-bg^3
\end{equation}
for calculable numbers $a$ and $b$ as suggested by (\ref{fine1}) and the standard result in the absence of gravity. It can then be seen that there is a possible fixed point away from $g=0$. If we ignore the running of $\kappa$ and $\Lambda$, the fixed point occurs at $g_\star^2={a}\kappa^2\Lambda/{b}$ provided that the right hand side is positive. If we take $\Lambda>0$, as present observations indicate \cite{Wmap}, and take $a<0$ as found above, then we require $b<0$. The resulting fixed point will not be ultraviolet stable, and in the absence of gravity ($a=0$) would correspond to a theory that is not asymptotically free, Yang-Mills theory with too many fermions for example. Thus it appears as if quantum gravity can alter the location of a fixed point. Ignoring the renormalization group evolution of $\kappa$ and $\Lambda$ results in a value of $g_\star$ that is tiny, due to the smallness of $\Lambda$, so the phenomenological importance of a putative fixed point is questionable. At this stage it is not clear if $a<0$ is a universal feature or is theory specific. It is still possible, as suggested in \cite{RobWilczek}, that quantum gravity corrections can render a theory that is not asymptotically free in the absence of gravity asymptotically free when gravity  is quantized, as we found for quantum electrodynamics. Whether or not this is the case, and the phenomenological relevance is worthy of further investigation. A puzzling aspect of the result obtained is that the cosmological constant is normally thought of as affecting gravity only at very large distances, whereas our result shows that there can be an effect at small distances. It is possible that running gauge coupling constants could give useful information about the cosmological constant, establishing an intriguing connection between macroscopic and microscopic realms.

{\bf Acknowledgements:} I would like to thank A. Rodigast for supplying \cite{RodigastPhD} and E. J. Copeland for supplying \cite{Wmap}. I am very grateful to the referees for suggesting many stylistic changes and for pointing out an error that affected the number multiplying the pole term in (\ref{gammadiv}).

\end{document}